\documentstyle[preprint,eqsecnum,aps]{revtex}

\begin{document}
\draft
\title{Inflation in Multidimensional Quantum Cosmology}
\author{Enrico Carugno$^{1}$, Marco Litterio$^{2}$\thanks{Corresponding
author. Electronic
address: litterio@astrom.astro.it},
Franco Occhionero$^1$, and
 Giuseppe Pollifrone$^{3}$\thanks{Electronic address:
pollifrone@roma1.infn.it}}
\address{$^{1}$ Osservatorio Astronomico di Roma,
via del Parco Mellini 84, 00136 Roma, Italy\\
$^{2}$ Istituto Astronomico, Universit\`a ``La Sapienza'',\\
 Via G. M. Lancisi 29, 00161 Roma, Italy\\
$^{3}$ Dipartimento di Fisica, Universit\`a
di Roma ``La Sapienza", and \\
INFN, Sezione di Roma,
Piazzale Aldo Moro 2, 00185 Roma, Italy}
\date{\today}
\maketitle
\begin{abstract}
We extend to multidimensional cosmology Vilenkin's prescription of tunnelling
from nothing for the quantum origin of the observable Universe.
Our model consists of a $D+4$-dimensional spacetime of topology 
${\cal R}\times {\cal S}^3 \times{\cal S}^D$, with a 
 scalar
field (``chaotic
inflaton'') for the matter component. Einstein 
 gravity and Casimir
compactification are assumed. The resulting minisuperspace is
3--dimensional.
 Patchwise we find an approximate analytic solution of the
Wheeler--DeWitt
 equation through which we discuss the tunnelling picture
and the probability 
 of  nucleation of the
 classical Universe with
compactifying extra dimensions.
Our conclusion is that the most likely initial conditions, although they
do not lead to the compactification of the internal space, still yield
(power-law) inflation for the outer space. The scenario is physically
acceptable because the inner space growth is limited to $\sim 10^{11}$ in
100 e-foldings of inflation, starting from the Planck scale.
\end{abstract}
\section{\bf Introduction}
\label{sec:introd}

A multidimensional spacetime is often required by fundamental physics
\cite{Kolb}. From the cosmological point of view, it is interesting to
exploit
 the phenomenology related to the extra space--like dimensions (the
``internal
 space''). Since such extra dimensions have not left any trace in
the observable
 Universe from the nucleosynthesis onward, the question is
what has kept them
 small and (almost) static, {\em i.e.}, how to have a
stable compactification. A
 static solution has been provided  using an
effective potential with quantum
 effects \cite{Casimir} ({\it Casimir}
compactification scheme) or a condensed
 antisymmetric tensor \cite{FR}
({\it monopole} compactification scheme). Later
 on, the effective action
for quantum effects and its low temperature limit have
 been evaluated
\cite{Yosh} for more general, non static, backgrounds, of
 cosmological
interest. In a closed cosmological model characterized by two
 scale
factors, $a(t)$ associated to the ordinary 3--dimensional space and
 $b(t)$
to the extra dimensions, the general formula for the effective potential
assumes a simple expression  in the flat--space limit \cite{Yosh} $a(t)\gg
b(t)$. This is the Casimir potential commonly used in literature, which we
will
 make use of in this paper. 

The scale factor $b(t)$ can be treated as a scalar field, the {\it
dilaton},
 in a 4--dimensional spacetime, under the action of the Casimir
potential. It
 has been shown \cite{Sokol} that for the dilaton field to be
an ordinary field,
 {\em i.e.}, for it to have a canonical kinetic term, an
appropriate and unique
 field redefinition is required. This point is often
neglected in the current
 literature, but at the price of an intrinsecal
instability \cite{Sokgol}. 

Once the dynamics of the internal space is reduced to that of an ordinary
scalar field in four dimensions, one may ask whether the dilaton can drive
inflation \cite{Kolb} in external space. The answer has been worked out in
the
 Casimir \cite{Okayosh} and in the monopole \cite{Hall} model. In both
cases the
 theory includes a cosmological constant. The resulting
potentials,
 qualitatively very similar, have a local maximum that, for
specific, fine
 tuned, initial conditions, makes possible some sort of
inflation. But, as
 Halliwell \cite{Hall} has shown explicitely for the
monopole case, and as we
 will discuss later for the Casimir case, the top
of the potential is not flat
 enough for the duration of the inflationary
period to be satisfactory. Thus we
 are forced to introduce another scalar
field, the {\it inflaton}, with its own
 appropriate potential. It has been
shown \cite{Amend}, indeed, that in this
 way, it is possible to have an
inflationary stage, without, in the meantime,
 destroying compactification,
provided suitable initial conditions are given. 
 Let us go through this
specific model in some detail since it will be adopted
 also in the present
paper. 

  The model \cite{Amend} consists of:
\begin{itemize}
\item
an $N$--dimensional universe of topology
${\cal R}\times{\cal S}^3\times{\cal S}^D$, with $N=D+4$;
\item
quantum corrections of the Casimir type, {\em i.e.}, one--loop vacuum
fluctuations of matter fields in a compact space;
\item
a scalar field $\phi$ driving inflation;
\item
a multidimensional cosmological constant tuned to ensure that no effective
4--dimensional cosmological constant  appears.
\end{itemize}

After dimensional reduction \cite{dimred}  and field
redefinition \cite{Sokol}, the internal space can be treated as
an ordinary scalar field $\sigma$ in Einstein theory.
The whole dynamics---standard $4$--dimensional model
with two scalar fields, $\sigma$ (dilaton) and $\phi$
(inflaton)---is then derived from the action:
\begin{eqnarray}
\label{finaction}
S & = & \int d^4\!x \, \sqrt{-g}\, \left[-\frac{1}{16\pi G_N}
 R+\frac{1}{2}g^{\mu\nu}\partial_{\mu}
\sigma\partial_{\nu}\sigma\right.\nonumber\\
&&\left.+\frac{1}{2}g^{\mu\nu}\partial_\mu\phi
	\partial_\nu\phi -U(\sigma,\phi)\right]\,\,.
\end{eqnarray}
In (\ref{finaction}), $R$ is the $4$--dimensional scalar curvature and
\begin{equation}\label{potent}
U(\sigma,\phi)=V_0(\sigma)+e^{-D\sigma/\sigma_0}\,V_1(\phi)
\end{equation}
is the combined potential for the two scalar fields. The $\sigma$ field is
related to the scalar factor of the internal space $b$ through
\begin{equation}\label{confor}
\sigma=\sigma_0 \ln\frac{b}{b_0}\,\,,\qquad{\rm where}\qquad
\sigma_0\equiv\sqrt\frac{D(D+2)}{16\pi G_N}\,\,,
\end{equation}
and  $b_0$ is the expected present size of the  radius of the internal
space, corresponding to the static ground state $\sigma=0$. Such a state
exists if one assumes  a potential $V_0(\sigma)$ of the Casimir type:
\begin{eqnarray}
\label{Casimpot}
V_0(\sigma) = \frac{(D-1)\sigma_0^2}{(D+4)b_0^2}
&&\left[{2\over D+2}e^{-2(D+2)\sigma/\sigma_0}
+e^{-D\sigma/\sigma_0}\right.\nonumber\\
&&\left.-{D+4\over D+2}e^{-(D+2)\sigma/\sigma_0}\right]\,\,.
\end{eqnarray}
The $\phi$ field can drive the inflation of external space if we assume
an appropriate potential $V_1(\phi)$. For example, in the chaotic inflation
scenario, $\phi$ is a ``classical'' field with the U--shaped
potential:
\begin{equation}\label{chaopot}
V_1(\phi)=\lambda \frac{\phi^4}{4}\,\,.
\end{equation}
Then the (total) potential (\ref{potent}) is shown in Fig. 2.
For sufficiently small initial values of
$|\phi|$, there exists a channel of stability where the
classical universe can undergo an inflationary stage in external
space while remaining compactified in internal space (``slow rolling''
of the inflaton along the potential in the $\phi$--direction).
The $\phi$ field ends its evolution with damped oscillations around its ground
state $\phi=0$, leading the universe to the present Friedmannian stage,
if one assumes $b_0<10^5\sqrt{D}l_{Pl}$.

In this paper, we look for inflationary solutions, with particular regard
to the scenario described above \cite{Amend},
in the framework of quantum cosmology \cite{QC1}.
In fact, through the wave function of the Universe $\Psi$ we can evaluate the
probability of different initial conditions (although this evaluation is a
problem in itself, as we will discuss with some more detail in
Section~\ref{sec:iiib}).
To find a specific $\Psi$ we need solving the Wheeler--DeWitt equation with a
given law of boundary conditions.
Two prescriptions are commonly used in literature:
 the ``no--boundary'' conditions proposed by Hartle and
Hawking \cite{QC2},  and
Vilenkin's ``tunnelling from nothing'' \cite{Vile1,Vile2}.
The former have been more widely used, but the latter usually
give a larger measure of classical solutions with sufficient
inflation \cite{QC1,Carugno}.

Therefore (see also \cite{Kamen}) we will consider only Vilenkin conditions
in this paper. They can be stated as follows:
\begin{itemize}
\item take only the outgoing modes of the wave function at the
singular boundary of the superspace;
\item impose a finiteness condition on the wave function.
\end{itemize}
In quantum cosmology the superspace is the configuration space of the universe
(3--geometries, local configurations of matter fields): the classical spacetime
corresponds to the region of the superspace in which the wave function
oscillates with large phase values. The non--singular boundary of the
superspace is that part of the boundary that includes 3--geometries given
through a slicing of a regular 4--geometry: the rest of the boundary is called
``singular''.

In many one-- or two--dimensional minisuperspace models \cite{QC1} Vilenkin
conditions describe a classical universe that nucleates via a tunnelling from
the non--singular boundary of the minisuperspace (the {\em nothing}) through
the superpotential barrier. Although such a picture does not always hold,  the
Vilenkin wave function still  selects  inflationary initial conditions for a
wide class of models, in $4$--dimensional cosmology.  In particular, it has
success \cite{Vile2} in the chaotic inflation scenario (Einstein gravity plus
chaotic inflaton). In this paper we extend Vilenkin's idea to multidimensional
cosmology, introducing\footnote{Quantum cosmology
with Hartle--Hawking boundary conditions for a model Einstein gravity plus
dilaton was studied by Okada and Yoshimura \cite{Okayosh}. They also
very briefly  comment that the Vilenkin boundary conditions
predict that the nucleation of a classical universe like our one would
be exponentially suppressed. As we mentioned above, in such a model,
the inflationary scenario is not satisfying already at the classical level.
Thus it is necessary to introduce the inflaton $\phi$.}
 in this chaotic scenario the dynamics of the dilaton.

In Section \ref{sec:ii}, we derive the Wheeler--DeWitt equation for our model.
The resulting minisuperspace is 3--dimensional, which makes the analysis very
difficult. We evaluate the wave function, in the semiclassical limit, in all
its interesting subregions. In Section \ref{sec:iii} we comment about the
tunnelling picture and the meaning of nothing in multidimensional cosmology, we
discuss the inflationary solutions and the quantum stability of the internal
space, and we present our conclusions.

\section{\bf The wave function}
\label{sec:ii}
\subsection{\bf The Wheeler--DeWitt equation}
\label{sec:iia}

We assume that the external space has the metric of a 3--sphere of radius
$a$, and the internal space that of a $D$--sphere of radius $b$.
Let us express all the space--time coordinates in units of
$\sqrt{2G_N/3\pi}$, and let us introduce the dimensionless fields
\begin{equation}\label{sig}
\sigma^{New}=\sqrt{\frac{4\pi G_N}{3}}\sigma\,\,,\qquad
\phi^{New}=\sqrt{\frac{4\pi G_N}{3}}\phi\,\,;
\end{equation}
at the same time we drop the superscript
$New$. In our model the metric of the world
``4--dimensional space--time + $\sigma,\phi$ fields'' is
\begin{equation}\label{metric}
ds^2=\frac{2G_N}{3\pi}\left(N^2dt^2-a^2(t)d\Omega_3^2\right)\,\,,
\end{equation}
After a simple integration,
the action (\ref{finaction}) becomes
\begin{equation}\label{homact}
S=\int dt \frac{N}{2}\left[-\frac{a\dot a^2}{N^2}+\frac{a^3}{N^2}\left(
\dot\sigma^2+\dot\phi^2\right)+a-a^3U(\sigma,\phi)\right]\,\,,
\end{equation}
where a dot denotes time derivatives.
The spatial degrees of freedom of the inflaton
have also been frozen in the minisuperspace scheme. The potential
in (\ref{homact}) is now
\begin{eqnarray}\label{finpot}
U(\sigma,\phi)&=&V_D(\sigma)+e^{-D\sigma/\sigma_D}\,V(\phi)\nonumber\\
V_D(\sigma) &=&K\left[{2\over D+2}e^{-2(D+2)\sigma/\sigma_D}
+e^{-D\sigma/\sigma_D}\right.\nonumber\\
&&~~~~~~~\left.-{D+4\over D+2}
e^{-(D+2)\sigma/\sigma_D}\right]\,\,,\nonumber\\
V(\phi)&=&\lambda \frac{\phi^4}{4}\,\,,\nonumber\\
K&=&\frac{2(D-1)\sigma_D^2}{(D+4)b_0^2}\,\,,\quad
\sigma_D=\sqrt{\frac{D(D+2)}{12}}\,\,.
\end{eqnarray}
The Wheleer--DeWitt equation in the minisuperspace of coordinates
$a,\sigma,\phi$, is then
\begin{equation}\label{wdw}
\left[\frac{1}{a}\partial_a^2-\frac{1}{a^3}\left(\partial_{\sigma}^2+
\partial_{\phi}^2\right)-w(a,\sigma,\phi)\right]\Psi(a,\sigma,\phi)=0\,\,,
\end{equation}
where
\begin{equation}\label{sup0}
w(a,\sigma,\phi)=a\left[1-a^2U(\sigma,\phi)\right]\,\,,
\end{equation}
is the superpotential of the universe.

Since we want to find the most probable initial conditions for the classical
motion of the universe, we consider the semiclassical
wave function ({\em i.e.}, the lowest order in the WKB expansion
\cite{QC1}).
In (\ref{wdw}), we assumed a simple factor ordering in the superhamiltonian:
in the semiclassical limit the arbitrariness of the choice of the factor
ordering does not affect the solution.
The equation
\begin{equation}\label{star}
a=a_{\star}(\sigma,\phi)=\frac{1}{\sqrt{U(\sigma,\phi)}}
\end{equation}
defines a surface of constant superpotential $w=0$ in the minisuperspace.  Eq.
(\ref{sup0}) describes a superpotential barrier in the $a$--direction: Eq.
(\ref{star}) separates the region $0<a<a_{\star}(\sigma,\phi)$, below the
barrier, from the region $a>a_{\star}(\sigma,\phi)$, beyond the barrier. The
formal analogy between the Wheeler--DeWitt  equation and a ``zero energy''
Schroedinger equation (at least when the kinetic contributions of the matter
fields are negligible) and this structure of the superpotential suggest that
the universe can nucleate at the (presumed) classical/quantum boundary
$a=a_{\star}(\sigma,\phi)$ through a quantum tunnelling process from the
configuration $a=0$. To investigate this possibility, we have to evaluate the
Vilenkin wave function. We will look for the approximate analytic solution of
the Wheeler--DeWitt  equations in the relevant regions of the minisuperspace.
In this section we will proceed as follows. First we show that a behaviour of
``nothing state'' for the wave function is present under the barrier. Then we
split the minisuperspace in two regions: i) the region of {\em small} $|\phi|$,
{\em i.e.}, where the term with $V(\phi)$ in the potential is negligible; ii)
the region of {\em large} $|\phi|$ where, on the contrary, the term with
$V(\phi)$ is dominant. We evaluate the Vilenkin wave function in these regions
in the WKB limit. Finally, we confirm our results using the method of the
constant $w$ surfaces, developed by Halliwell \cite{Hall}.

\subsection{\bf The solution of nothing}
\label{sec:iib}

We now look for the wave function in the region where $a$ is {\em small},
{\em i.e.}, under the barrier of superpotential:
\begin{equation}\label{small}
a^2\ll a^2_{\star}(\sigma,\phi)\,\,.
\end{equation}
The Wheeler--DeWitt  equation (\ref{wdw}) reduces locally to
\begin{equation}\label{wdw2}
\left(a^2\partial_a^2-\partial_{\sigma}^2-\partial_{\phi}^2-a^4\right)\Psi=0
\,\,.
\end{equation}
Then Eq. (\ref{wdw2}) is reducible, through the substitution $\Psi=
\Theta(a,\sigma)\Gamma(\phi)$,  to two decoupled equations in $\Theta$ and
$\Gamma$, parametrically depending on the separation constant $E$, whose value
can be determined by a matching of the solutions found in nearby regions. In
the following we adopt the heuristic method of Halliwell \cite{Hall}, who
neglects the separation constant with respect to the superpotential in the
regions of the minisuperspace where the latter is large in modulus. This is
equivalent to assume \cite{Vile3} that the wave function is asymptotically
$\phi$--independent, where the superpotential also has this  property. Indeed,
for $U(\sigma,\phi)\ll 1$, values of $a$ such that $a^2\gg 1$ also belong to
region (\ref{small}). For them, the superpotential in Eq.~(\ref{wdw2}) is, in
modulus, much greater than $1$. The separation constant can be neglected and we
can assume that, at the WKB lowest order, the wave function is proportional to
the solution $\Theta(a)$ of the equation
\begin{equation}\label{wdw3}
\left(d^2_a-a^2\right)\Theta(a)=0\,\,.
\end{equation}
Introducing the auxiliary variable $\Gamma(a)=\Theta(a)/a^{1/2}$
and the transformation
$v=a^2/2$, Eq.~(\ref{wdw3}) reduces to the modified Bessel equation
\begin{equation}\label{bess}
v^2d^2_v\Gamma+vd_v\Gamma-\left(v^2+\frac{1}{16}\right)\Gamma=0\,\,,
\end{equation}
whose independent solutions are the well known modified Bessel functions
of order $1/4$, $I_{1/4}(v)$ and $K_{1/4}(v)$. Going back to the old variables,
we find the growing solution $a^{1/2}I_{1/4}(a^2/2)$ and the decreasing
solution
$a^{1/2}K_{1/4}(a^2/2)$ in the $a$--direction.
To select one of them, we impose a matching condition with the solution
(\ref{A}),
discussed in the following subsection, that holds for small $|\phi|$.
Considering only the dominant exponential factors for $a\gg 1$,  the matching
 in the
intersection of region (\ref{cond1}) with region
(\ref{small}), gives
\begin{equation}\label{noth}
\Psi=a^{\frac{1}{2}}K_{1/4}(\frac{a^2}{2})\,\,,
\end{equation}
solution that satisfies locally the Vilenkin boundary condition.
Eq.~(\ref{noth}) is the well known solution of nothing \cite{Vile1}: it has
been found by Vilenkin in the limit of small $a$ in the 4--dimensional model
with topology ${\cal R}\times{\cal S}^3$ and inflaton, without dilaton. This
wave function is monotonic, does not depend on the matter fields, and it has a
peak, around $a=0$, whose width is of the order of the Planck length.

Eq.(\ref{noth}) is the wanted solution of nothing. Nevertheless, because of
solution (\ref{D}), the picture of tunnelling from nothing through the
superpotential barrier cannot be straightforwardly extended to our
multidimensional model. We will discuss in detail this problem in Section
\ref{sec:iiia}.

\subsection{\bf Nucleation of the classical Universe?}
\label{sec:iic}
\subsubsection{\bf The minisuperspace region of small $|\phi|$}

In this case the condition
\begin{equation}\label{cond1}
V(\phi)\ll e^{D\sigma/\sigma_D}V_D(\sigma)
\end{equation}
holds. The Wheeler--DeWitt   equation can be rewritten as
\begin{equation}\label{asy}
\left\{a^2\partial_a^2-\partial_{\sigma}^2-
\partial_{\phi}^2-a^4\left[1-a^2V_D(\sigma)\right]\right\}\Psi(a,\sigma,
\phi)=0\,\,.
\end{equation}
Then Eq. (\ref{asy}) is reducible, through the substitution $\Psi=
\Theta(a,\sigma)\Gamma(\phi)$,  to two decoupled equations in $\Theta$ and
$\Gamma$, parametrically depending on the separation constant $E$. Following
again the heuristic method of Halliwell \cite{Hall}, of neglecting the
separation constant with respect to the superpotential in the regions of the
minisuperspace where the latter  is large in modulus, to the WKB lowest order,
$\Psi$ is then proportional to the solution of equation
\begin{equation}\label{asy2}
\left\{a^2\partial_a^2-\partial_{\sigma}^2
-a^4\left[1-a^2V_D(\sigma)\right]\right\}\Theta(a,\sigma)=0\,\,.
\end{equation}

The solution of Eq.(\ref{asy2}) with Vilenkin boundary conditions
is well known \cite{Vile2} in the region of the minisuperspace where the
potential $V_D(\sigma)$ is  ``sufficiently flat,'' {\em i.e.}, where
\begin{equation}\label{cond2}
\left|\frac{dV_D(\sigma)}{d\sigma}\right|\ll {\rm max}\left\{V_D(\sigma),
\frac{1}{a^2}\right\}\,\,.
\end{equation}
Such a solution, to the lowest order of the WKB expansion, is
\begin{equation}\label{A}
\Theta(a,\sigma)=\left\{\begin{array}{cc}
\exp\left\{-\left[1-\left(1-V_D(\sigma) a^2\right)^{3/2}\right]/[3V_D(\sigma)]
\right\} &  \left(a^2V_D(\sigma)<1\right)\\
		&   \\
\exp\left\{-\left[1+i\left(V_D(\sigma) a^2-1\right)^{3/2}\right]/[3V_D(\sigma)]
	\right\}& \left(a^2V_D(\sigma)>1\right)
	\end{array}\right.
\end{equation}
In our model, nevertheless, unlike in Ref.~\cite{Vile2}, the potential
$V_D(\sigma)$ has a
strongly asymmetric form for $\sigma>0$ and $\sigma<0$. Then condition
(\ref{cond2}) and solutions (\ref{A}) do not hold in an important region of
the minisuperspace (Fig. 3): the region $a^2V_D(\sigma)>1$, $\sigma<0$.
For $\sigma\to -\infty$, in particular, the barrier of superpotential
becomes narrow with respect to the configuration $a=0$: thus, this
region is particularly interesting for tunnelling conditions.
To find out the wave function here, we adopt the following procedure.
Under the transformation $\alpha=\log a$, the Wheeler--DeWitt  equation
(\ref{wdw}) becomes
\begin{equation}\label{asy3}
\left\{\partial_{\alpha}^2-\partial_{\alpha}-\partial_{\sigma}^2-
\partial_{\phi}^2-e^{4\alpha}\left[1-e^{2\alpha}U(\sigma,\phi)\right]\right\}
\Psi(\alpha,\sigma,\phi)=0\,\,.
\end{equation}
In the semiclassical limit, we can omit the first derivative: this
 is equivalent to a particular choice \cite{Hall} of the factor ordering
in the Wheeler--DeWitt  equation, to which, in the
semiclassical limit, the wave function is insensitive. The
Wheeler--DeWitt  equation  is now
\begin{equation}\label{asy4}
\left\{\begin{array}{l}\left[\partial_{\alpha}^2-\partial_{\sigma}^2-
\partial_{\phi}^2-W(\alpha,\sigma,\phi)\right]
\Psi(\alpha,\sigma,\phi)=0\\
W(\alpha,\sigma,\phi)=e^{4\alpha}\left[1-e^{2\alpha}U(\sigma,\phi)\right]
\end{array}\right.\,\,.
\end{equation}
Repeating the preceding
discussion, see Eq. (\ref{asy2}), we find that $\Psi$ is proportional to the
solution $\Theta$ of the equation
\begin{equation}\label{asy5}
\left\{\partial_{\alpha}^2-\partial_{\sigma}^2-
e^{4\alpha}\left[1-e^{2\alpha}V_D(\sigma)\right]\right\}
\Theta(\alpha,\sigma)=0\,\,.
\end{equation}
For
\begin{equation}\label{region2}
a^2V_D(\sigma)\gg 1\,\,,\qquad \sigma\ll -\sigma_D/D\,\,,
\end{equation}
 Eq.(\ref{asy5}) becomes
\begin{equation}\label{asy6}
\left[\partial_{\alpha}^2-\partial_{\sigma}^2+\frac{2K}{D+2}
\exp\left(6\alpha-2\frac{D+2}{\sigma_D}\sigma\right)
\right]
\Theta(\alpha,\sigma)=0\,\,.
\end{equation}
Under the rotation
\begin{eqnarray}\label{rota}
 \tilde\alpha & = & \frac{1}{g_D}\left(-\frac{D+2}{3\sigma_D}
\alpha+\sigma\right)\nonumber\\
\tilde\sigma&=&\frac{1}{g_D}\left(\alpha-\frac{D+2}{3\sigma_D}\sigma\right)
\nonumber\\
g_D &=&\{[(D+2)/3\sigma_D]^2-1\}^{1/2}
\end{eqnarray}
(for $D$ positive integer, $g_D$ is
always a positive number smaller than 1),  Eq.(\ref{asy6}) becomes separable:
\begin{equation}\label{asy7}
\left[\partial_{\tilde{\alpha}}^2-\partial_{\tilde{\sigma}}^2+\frac{2K}{D+2}
\exp\left(6\tilde{\sigma}g_D\right)\right]
\Theta(\tilde{\alpha},\tilde{\sigma})=0\,\,.
\end{equation}
Again the contribution of the separation constant
is negligible where the superpotential becomes sufficiently large.  This is
the  region where
 $\tilde{\sigma}$ is large  ({\em i.e.}, condition (\ref{region2}) holds true).
Proceeding as before $\Theta$  turns out to be asymptotically proportional
to the solution
$\Omega(\tilde{\sigma})$ of
the equation
\begin{equation}\label{asy8}
\left[-d_{\tilde{\sigma}}^2+\frac{2K}{D+2}
\exp\left(6\tilde{\sigma}g_D\right)\right]
\Omega(\tilde{\sigma})=0\,\,.
\end{equation}
Eq.(\ref{asy8}) admits monotonic independent solutions that can
be written (in the original variables $a ,\sigma$) as
\begin{equation}\label{D}
\Omega=\exp\left\{\pm\sqrt{\frac{2K}{D+2}}\frac{a^3}{3g_D}
e^{-(D+2)\sigma/\sigma_D}\right\}\,\,.
\end{equation}
Vilenkin's regularity condition $|\Psi|<\infty$ selects the decreasing mode of
 $\Omega$ in the region $a\to\infty$ and/or $\sigma\to -\infty$ of the singular
boundary.

Note that, with an analogous method, it is possible to derive
the solution of Eq.(\ref{asy2}) in the region
\begin{equation}\label{region3}
a^2V_D(\sigma)\gg 1\,\,,\qquad \sigma\gg\sigma_D/D\,\,,
\end{equation}
where again Eqs. (\ref{cond2}), (\ref{A}) do not hold.
Here the Wheeler--DeWitt  equation becomes
\begin{equation}\label{asi9}
\left[\partial_{\alpha}^2-\partial_{\sigma}^2+K
\exp\left(6\alpha-\frac{D\sigma}{\sigma_D}\right)
\right]
\Theta(\alpha,\sigma)=0\,\,,
\end{equation}
from which, under the rotation
\begin{eqnarray}\label{rota2}
 \tilde\alpha & = & \frac{1}{g_D}\left(\alpha-
\frac{D\sigma}{6\sigma_D}\right)\nonumber\\
		\tilde\sigma & = & \frac{1}{g_D}\left(-
\frac{D\alpha}{6\sigma_D}+\sigma\right)\nonumber\\
g_D & = & \left[1-(D/6\sigma_D)^2\right]^{1/2}
\end{eqnarray}
(for $D$ positive integer, $g_D$ is always real, positive and smaller than 1)
we get
\begin{equation}\label{asy10}
\left[\partial_{\tilde{\alpha}}^2-\partial_{\tilde{\sigma}}^2+K
e^{\left(6g_D\tilde{\alpha}\right)}\right]
\Theta(\tilde{\alpha},\tilde{\sigma})=0\,\,.
\end{equation}
For large $\tilde{\alpha}$, {\em i.e.}, in the region (\ref{region3}), Eq.
(\ref{asy10}) admits independent solutions that, in the original variables, can
be written as
\begin{equation}\label{A2}
\Theta=\exp\left(\pm i\frac{\sqrt{K}}{3g_D} a^3
e^{-D\sigma/(2\sigma_D)}\right)\,\,.
\end{equation}
Of the two oscillating modes (\ref{A2}), Vilenkin's boundary conditions select
only the outgoing one ({\it i.e.}, the second in (\ref{A2})) at the singular
boundary
of the minisuperspace. Note that in the region (\ref{region3}), the amplitude
of the oscillations of the wave function changes slowly: contrary to
what happens in the strip defined by $a^2V_D>1$ and $\sigma\sim
\sigma_{\rm top}$ [see (\ref{cond2}) and (\ref{A}); $\sigma_{\rm top}$
is the value of $\sigma$ for which $V_D(\sigma)$ has a local maximum],
here the wave function is not strongly peaked.

\subsubsection{\bf The minisuperspace region of large $|\phi|$}

In this case
\begin{equation}\label{cond1b}
V(\phi)\gg e^{D\sigma/\sigma_D}V_D(\sigma)\,\,.
\end{equation}
For {\em large} $a$,
\begin{equation}\label{large}
a^2\gg a^2_{\star}(\sigma,\phi)\,\,,
\end{equation}
Eq.(\ref{asy4}) reduces to
\begin{equation}\label{notil}
\left[\partial_{\alpha}^2-\partial_{\sigma}^2-\partial_{\phi}^2+
\exp\left(6\alpha-\frac{D\sigma}{\sigma_D}\right)V(\phi)\right]
\Psi=0\,\,.
\end{equation}
Through the rotation (\ref{rota2}), Eq.(\ref{notil}) can be written
\begin{equation}\label{wdw4}
\left[\partial_{\tilde{\alpha}}^2-\partial_{\tilde{\sigma}}^2-\partial_{\phi}^2
+e^{\left(6g_D\tilde{\alpha}\right)}V(\phi)\right]
\Psi=0\,\,.
\end{equation}
Once again, since the superpotential of Eq.(\ref{wdw4}) does not depend on
$\tilde{\sigma}$, we can disregard the $\tilde{\sigma}$--dependence of the
wave function where the superpotential is large in modulus, {\em i.e.}, in the
region (\ref{cond1b}) and (\ref{large}). The wave function is thus proportional
to the solution $\Sigma(\tilde{\alpha},\phi)$ of equation
\begin{equation}\label{wdw5}
\left[\partial_{\tilde{\alpha}}^2-\partial_{\phi}^2
+e^{\left(6g_D\tilde{\alpha}\right)}V(\phi)\right]
\Sigma=0\,\,.
\end{equation}
Under the transformation
\begin{eqnarray}\label{tran}
\tilde{\alpha}&=&\frac{1}{g_D}\log(g_D q)\nonumber\\
\phi&=&\frac{1}{g_D}x\,\,,
\end{eqnarray}
Eq.(\ref{wdw5}) becomes
\begin{equation}\label{wdw6}
\left[\partial_q^2+\frac{1}{q}\partial_q-\frac{1}{q^2}\partial_x^2+q^4V(x)
\right]\Sigma=0\,\,.
\end{equation}
For $q^2V(x)\gg 1$ (and with a factor
ordering $p=1$ in the notation of Ref.~\cite{Vile2},
Eq.(\ref{wdw6}) is formally identical to a Wheeler--DeWitt  equation for
which the Vilenkin wave function is already known. Under the further limit
\begin{equation}\label{limit2}
\left|\frac{1}{V(x)}\frac{dV}{dx}\right|\ll 1\,\,,
\end{equation}
to the lowest WKB order,
\begin{equation}\label{sigh}
\Sigma(q,x)=\exp\left[-\frac{1}{3V(x)}\right]\exp\left[-i\frac{q^3}{3}
V^{\frac{1}{2}}(x)\right]\,\,.
\end{equation}
Going back to the original variables, the wave function can be written
\begin{equation}\label{sol}
\Psi= \exp\left[-\frac{1}{3g_D^4V(\phi)}\right]\exp\left[-i
\frac{a^3}{3g_D}e^{-D\sigma/2\sigma_D}V^{\frac{1}{2}}(\phi)\right]\,\,,
\end{equation}
in the region of intersection of (\ref{cond1b}), (\ref{large}) and
\begin{eqnarray}\label{cond1c}
a^2\gg a^2_{\star}(\sigma,\phi)e^{-2D\sigma/3\sigma_D}g_D^2\,\,,\nonumber\\
\left|\frac{1}{V(\phi)}\frac{dV}{d\phi}\right|\ll 1\,\,.
\end{eqnarray}
Note that in the absence of the dilaton ($\sigma=0$) and of extra
dimensions ($D=0$, {\em i.e.}, $g_D=1$) solution (\ref{sol}) reduces just to
the Vilenkin solution for the model ${\cal R}\times{\cal S}^3$ with
inflaton \cite{Vile2}.

\subsection{\bf The method of null surfaces}
\label{sec:iid}
It is possible to confirm qualitatively the structure of the wave function
found in the different regions of the minisuperspace, using the method of the
surfaces of constant superpotential \cite{Hall}. In our model
 the null surfaces of constant superpotential have equation
\begin{equation}\label{null}
a^2=\frac{4}{6U\pm\sqrt{(\partial_{\sigma}U)^2+(\partial_{\phi}U)^2}}\,\,,
\end{equation}
while the surfaces of null superpotential have equation
\begin{equation}\label{sup}
a^2=\frac{1}{U}\,\,,
\end{equation}
and $a=0$. The surfaces of equation (\ref{null}) separate the regions of the
minisuperspace where the surfaces of constant superpotential are of opposite
kind (time-- or space--like). The surfaces of equation (\ref{sup}) separate the
regions of the minisuperspace where the surfaces of constant superpotential are
of opposite sign. Sufficiently far from the surfaces of null superpotential,
the local comparison between the sign and the kind of the surfaces of constant
superpotential allows us to understand qualitatively the  behaviour, whether
monotonic or oscillating, of the wave function (Table 1). Eqs.(\ref{null}) and
(\ref{sup}) confirm the structure of the wave function evaluated explicitly in
this section.

In particular, \underline{in the region of small $|\phi|$},
 (\ref{null}) reduces to
\begin{equation}\label{null2}
a^2=\frac{4}{6V_D(\sigma)\pm d_{\sigma}V_D(\sigma)}\,\,,
\end{equation}
and (\ref{sup}) reduces to
\begin{equation}\label{sup2}
a^2=\frac{1}{V_D(\sigma)}\,\,.
\end{equation}

Eqs.(\ref{null2}) and (\ref{sup2}) formally coincide with the expressions found
by Halliwell \cite{Hall} for a model with $D=2$ extra dimensions, with monopole
potential $V_D(\sigma)$ and without inflaton $\phi$. Through them (Figs.3 and
4), we find out that in the main part of the region $a>a_{\star}(\sigma,\phi)$,
$\sigma>0$, the wave function is oscillating (not only in the strip $\sigma\sim
\sigma_{\rm top}$ and for $\sigma\gg \sigma_D/D$, in the region
$a>a_{\star}(\sigma,\phi)$,  where it has been explicitly derived); it is
monotonic in the remaining part (not only for $a<a_{\star}(\sigma,\phi)$ or
$a>a_{\star}(\sigma,\phi)$, $\sigma\ll-\sigma_D/D$).

\underline{In the region
of large $|\phi|$} (\ref{cond1b}), provided $|V^{-1}dV/d{\phi}|\ll 1$,
Eq.(\ref{null}) becomes
\begin{equation}\label{null3}
a^2=\frac{2e^{D\sigma/\sigma_D}}{3V(\phi)\left[1\pm D/6\sigma_D\right]}\,\,,
\end{equation}
and (\ref{sup}) reduces to
\begin{equation}\label{sup3}
a^2=\frac{e^{D\sigma/\sigma_D}}{V(\phi)}\,\,.
\end{equation}
Since for any $D$ integer and positive $0.3<D/6\sigma_D <0.6$, such a result
confirms (Fig.5) the oscillating behaviour (\ref{sol}), beyond the
superpotential barrier ($a^2\gg a_{\star}^2(\sigma,\phi$)), and the monotonic
behaviour (\ref{noth}) below ($a^2\ll a_{\star}^2(\sigma,\phi$)). Thus in such
a region of the minisuperspace, the superpotential barrier defines, roughly
speaking, the classically allowed and the classically forbidden regions. For
$\sigma$ negative and  large, outside the region of large $|\phi|$,
(\ref{cond1b}), this is no longer true.

\section{\bf Discussion}
\label{sec:iii}
\subsection{\bf Tunnelling picture in multidimensional cosmology}
\label{sec:iiia}

The picture of tunnelling from nothing through a superpotential barrier has
been criticized \cite{Jun2} or rephrased \cite{Carugno,Vile3}. We point out
some new features thereof that were not included in its original formulation.

First, the very definition of nothing is not clear when treating
multidimensional space--times. In $4$--dimensional cosmology, nothing is the
non--singular boundary of the superspace, {\em i.e.}, that part of the boundary
of the superspace that includes $3$--geometries given through a slicing of a
regular $4$--geometry \cite{Vile1,Vile2}. In the equivalent $4$--dimensional
model, the extra dimensions play the role of a matter scalar field $\sigma$:
the non--singular boundary of the minisuperspace is then the configuration
$a=0$, $|\sigma|<\infty$, $|\phi|<\infty$. We will call it {\em external
nothing} since the internal space is assumed to be non--zero ($a=0$,
$0<b<\infty$,$|\phi|<\infty$). It is also possible to figure out a {\em total
nothing} defined as $a=b=0$, $|\phi|<\infty$. However, only external nothing is
acceptable when Casimir or monopole schemes are used. In fact, vacuum
fluctuations in the former scheme give a finite contribution only for $b\ne 0$;
analogously the antisymmetric tensor field introduced in the latter one is
regular on the internal $D$--sphere only for $b\ne 0$. In both cases, the
configurations with $b=0$  ({\em i.e.}, total nothing $a=b=0$, $|\phi|<\infty$
and {\em internal nothing} $0<a<\infty$, $b=0$, $|\phi|<\infty$) do not belong
to the non--singular boundary of the minisuperspace, but to the singular
one.\footnote{It is worth reminding that when one assumes the Hartle--Hawking
boundary conditions \cite{QC2} for the cosmic wave function in the path
integral approach, something analogous happens \cite{Hall}. Summing  over
compact Euclidean $4$--geometries and over matter fields configurations that
are regular over them, initial conditions $a=0,b\neq 0$ of the paths are
assumed.} External nothing is the only nothing configuration classically stable
because of the superpotential barrier (at least when the kinetic energy of the
matter fields is negligible) and over which the wave function is peaked. In our
model it is the best candidate for the tunnelling picture, but now a problem
arises with the tunnelling itself.

In quantum cosmology the tunnelling picture can meet troubles due to the
hyperbolic nature of the Wheeler--DeWitt equation. Already in
Ref.~\cite{Jun2}, it has been pointed out that in models such as non
minimally coupled scalar field and Bianchi type--IX, the superpotential barrier
can disappear, leaving the configuration of nothing exposed to the Lorentzian
($w<0$) region of the minisuperspace. This fact raises the question of whether
one could apply the Vilenkin boundary conditions at all. We have got a similar
problem, but in our case the barrier never disappears. In the quantum mechanics
analog of the tunnelling, one would expect a nucleation to be more likely where
the barrier becomes thinner. In our model, the barrier becomes narrow with
respect to the external nothing for  $U\to\infty$, {\em i.e.}, for both the two
configurations $\sigma\to -\infty$, $|\phi|  \le\infty$ and $|\sigma|<\infty$,
$|\phi|\to\infty$. Nevertheless beyond the barrier ({\em i.e.}, for
$a>a_{\star}(\sigma,\phi)$), $\Psi$ is monotonic for small $|\phi|$, $\sigma\ll
-\sigma_D/D$ (while it oscillates for large $|\phi|$). This proves that the
nucleation of the semiclassical universe in one of the two configurations
($\sigma\to-\infty$, $|\phi|\le\infty$) is not possible, while in the standard
tunnelling picture they should be equally probable. In this sense the
tunnelling through the superpotential barrier picture does not hold. We will
argue, following  Ref.~\cite{Vile3} (although, differently from here, there is
an exposed nothing in the Bianchi type--IX considered there), that this is not
a real problem. The tunnelling picture in quantum cosmology was born only as a
formal analogy between the Vilenkin cosmic wave function and the tunnelling
wave function of the usual quantum mechanics. This analogy is limited to a
finite class of models and to  particular regions of the superspace. Indeed the
superpotential barrier does not always separate classically allowed regions
({\em i.e.}, where $\Psi$ oscillates) from forbidden ones ({\em i.e.},
where $\Psi$ is monotonic) in the space of configurations, contrary to the
usual potential barrier in quantum mechanics. This happens because the kinetic
form of the gravitational superhamiltonian $H$ has a hyperbolic structure,
while in a usual quantum system it is elliptic. So, to changes in sign  of the
superpotential ({\em i.e.}, regions of the minisuperspace above and below the
barrier) does not necessarily correspond, in the classical constraint $H=0$,
the fact that real ``velocities'' become imaginary, or viceversa. The
tunnelling analogy can hold only when the kinetic form has a definite sign with
respect to the superpotential. In our model, this does not happen in one of the
two configurations where the superpotential barrier becomes narrow, because
here the kinetic energy of matter fields
is not negligible. That is to say that
$$\left(\hat p_{\sigma}^2+\hat p_{\phi}^2\right)\Psi=-\frac{1}{a^3}
\left(\partial_{\sigma}^2+\partial_{\phi}^2\right)\Psi\,\,,$$
gives a relevant contribution in the Wheeler--DeWitt  equation (\ref{wdw})
$\hat H\Psi=0$ ($\hat p$ operator of the canonical momenta). Nevertheless,
Vilenkin boundary conditions have their physical meaning
\cite{Vile1,Vile2,Carugno} that does not depend on the tunnelling  analogy.
Selecting only outgoing modes for the wave function at the singular
boundary,\footnote{More precisely, at that part of the singular boundary close
to where $\Psi$ oscillates.} they fix a ``time direction'' in
minisuperspace: the direction of this probability flux toward the boundary. In
the following of the paper we will refer to this {\em causal meaning} of
Vilenkin boundary conditions. We remark that the causal conditions of quantum
cosmology still select an inflationary scenario in 4--dimensional models
when the tunnelling analogy does not hold anymore: for example, in
anisotropic models \cite{Vile3}.

As a last comment, we note that since dimensional reduction \cite{dimred} is
not valid anymore for small $a$ ({\em i.e.}, where the flat--space limit $a\gg
b$ fails), the wave function might not have the form of the ``nothing
solution'' (\ref{noth}) in the region $a^2\ll a^2_{\star}(\sigma,\phi)$ of
minisuperspace. Nevertheless, as we have just shown, Vilenkin boundary
conditions can be applied also when the tunnelling picture fails, and the
semiclassical wave function is peaked (see Section \ref{sec:iiib}) on
asymptotic
classical solutions in the region $a\gg b$, where dimensional reduction holds.

\subsection{\bf Interpretation of the wave function.
	The classical limit.}
\label{sec:iiib}

The interpretation of the wave function of the universe is still matter of
debate. In this paper we will adopt a minimal  interpretative criterion. Let us
consider the semiclassical wave function: we will assume that only the strong
peaks of $|\Psi|^2$ select classical correlations among the dynamical variables
of the universe \cite{Carugno}. Weak variations of $|\Psi|^2$ do not select any
correlation. Such a criterion only needs the lowest WKB order and does not
require either the normalization or even the normalizability\footnote{An
alternative way to solve the problem of the non normalizability of the  wave
function is the introduction of conditional probabilities: see
Refs.~\cite{QC1,QC2}.}  of the wave function (for this reason in Section
\ref{sec:ii}, we omitted everywhere the pre--exponential WKB factors and the
normalization constants: note that the Vilenkin wave function derived above is
not normalizable). In literature other criteria to interpret the wave function
can be found. Since they introduce measures that are factorized to $|\Psi|^2$
in the semiclassical limit, it is reasonable to expect results qualitatively
similar \cite{QC1} in presence of strong peaks.

 The structure of the wave function found in the previous section is the
following. For {\em small} $a$ ({\em i.e.}, $a^2\ll a^2_{\star}(\sigma,\phi)$),
under the superpotential barrier, $\Psi$ has the asymptotic behaviour
(\ref{noth}) of Vilenkin nothing state. It is peaked on $a=0$,
monotonically decreasing for increasing $a$, independently of the matter fields
$\sigma,\phi$. For {\em large} $a$ ({\em i.e.},  $a^2\gg
a^2_{\star}(\sigma,\phi)$),  beyond the superpotential barrier, we recognize
two behaviours. \\ a) For {\em small} $|\phi|$ ({\em i.e.}, in the region
(\ref{cond1})), $\Psi$ is weakly dependent on $\phi$. This is due to the weak
local dependence on $\phi$ of $U(\sigma,\phi)$. $\Psi$ is monotonic for
$\sigma<0$, and oscillating for $\sigma>0$. The amplitude of the oscillations
has a peak around $\sigma=\sigma_{\rm top}$, where $V_D(\sigma)$ has a local
maximum. Here
\begin{equation}\label{peak1}
|\Psi|^2=\exp\left[-\frac{2}{3V_D(\sigma)}\right]\,\,.
\end{equation}
b) For {\em large} $|\phi|$ ({\em i.e.}, in the intersection of the regions
(\ref{cond1b}), (\ref{large}), (\ref{cond1c})), $\Psi$  oscillates. The
amplitude of the oscillations increases quickly  for
increasing $|\phi|$ (converging to a finite limit, at the lowest
WKB order). Here
\begin{equation}\label{tweenpeaks!}
|\Psi|^2=\exp\left[-\frac{2}{3g_D^4V(\phi)}\right]\,\,.
\end{equation}
So the amplitude of the oscillations of
$\Psi$, but not their phase, depends weakly on $\sigma$.

 To the lowest WKB order, the peaks in the oscillating structure of $\Psi$
corresponding to (\ref{peak1}) and (\ref{tweenpeaks!}) are $a$--independent.
They depend only on the potential respectively of the dilaton and of the
inflaton. (Note that $1>g_D^4>0.4$ when $0<D<\infty$). The potential of the
inflaton is unbounded in the region of large $|\phi|$.  The potential of the
dilaton is, on the contrary, limited around $\sigma=\sigma_{\rm top}$. However
$V_D(\sigma_{\rm top})\gg 1$ for values of the equilibrium radius smaller than
the Planck scale, $b_0\ll D^{1/2}$: in such a case the two peaks  of $\Psi$
could be of comparable amplitude. The quantitative comparison between the
probabilities of nucleation of the universe in the regions corresponding to the
two peaks is impossible without introducing conditional probabilities
\cite{QC1}. Nevertheless neither one of the two peaks (\ref{peak1}) and
(\ref{tweenpeaks!}) corresponds to a universe that is compatible with that
observable today. Let's analyze the classical evolution of a typical universe
selected by them.

In the semiclassical limit the cosmic wave function is
strongly peaked on the trajectories of superspace that satisfy the first
integral
\begin{equation}\label{int1}
p=\nabla F\,\,,
\end{equation}
of the classical equations,
where $F$ is the phase of the wave function, $p$ are classical canonical
momenta and $\nabla$ is the gradient on the superspace \cite{QC1}.

{}From (\ref{sol}) and (\ref{int1}), we find, in the region of large $|\phi|$,
the first integral
\begin{eqnarray}\label{int2}
\frac{\dot a}{a}&=&\frac{1}{g_D}\sqrt{V(\phi)}e^{-D\sigma/2\sigma_D}\,\,,
\nonumber\\
\dot{\sigma}&=&\frac{D}{6\sigma_Dg_D}\sqrt{V(\phi)}e^{-D\sigma/2\sigma_D}\,\,,
\nonumber\\
\dot{\phi}&=&-\frac{d_{\phi}V(\phi)}{6g_D\sqrt{V(\phi)}}
e^{-D\sigma/2\sigma_D}\,\,.
\end{eqnarray}
Here $|\Psi|^2$ selects just large $|\phi|$,
while it is weakly depending on $\sigma$: all the values of $\sigma$ in this
region are almost equally probable. Eqs.(\ref{int2}) are verified by an
ensemble
of classical solutions of ``velocity'' $\dot{\phi}/\phi<0$, $\dot{\sigma}>0$:
thus the typical universe selected by the causal conditions in this region
rolls down along the profile of the potential $U$ toward the configuration
$\sigma=+\infty$ ({\em i.e.}, $b=\infty$). In particular for $|\phi|\gg 1$,
we have $\dot{\sigma}^2\gg \dot{\phi}^2$, and in the region $|\phi|\gg 1$,
$\exp D\sigma/\sigma_D>V(\phi)$,
necessarily $\dot{\phi}^2$ gets asymptotically to zero.
 This happens because the potential of matter
fields is sufficiently flat locally only in the $\phi$--direction ({\em i.e.},
$|\partial_{\phi}U/U|\simeq|d_{\phi}V/V|\ll 1$). The typical classical universe
follows then a trajectory almost parallel to the $\sigma$ axis, that can be
written, sufficiently far from $\sigma_{in}$, as
\begin{eqnarray}\label{int3}
a&\propto & t^p\,\,, \qquad p=\frac{D+2}{D}\,\,,\nonumber\\
\sigma&\simeq &\frac{2\sigma_D}{D}\ln \left[\frac{D^2}{12g_D\sigma_D^2}
\sqrt{V(\phi_{in})}t\right]\,\,,\nonumber\\
\phi&\simeq &\phi_{in}\,\,,
\end{eqnarray}
(everywhere a subfixed $in$ indicates an initial value).
This is a power--law inflation \cite{Luc}, though rather weak for large
$D$. The classical universe corresponding to the peak of the semiclassical
function in the region of large $|\phi|$, undergoes an eternal inflation with
an unstable internal space (note that $a/b\propto t$).

The peak (\ref{peak1}) on the strip $\sigma\simeq \sigma_{\rm top}$ of the
region of small $|\phi|$ selects instead the first integral
\begin{eqnarray}\label{int4}
\frac{\dot a}{a}&=&\sqrt{V_D(\sigma_{\rm top})}\left[1-
\frac{\mu^2}{4V_D(\sigma_{\rm top})}(\sigma-
\sigma_{\rm top})^2\right]\,\,,\nonumber\\
\dot{\sigma}&=&\frac{\mu^2}{6\sqrt{V_D(\sigma_{\rm top})}}
(\sigma-\sigma_{\rm top})
\left[1+\frac{\mu^2}{4V_D(\sigma_{\rm top})}(\sigma-
\sigma_{\rm top})^2\right]\,\,,\nonumber\\
\dot{\phi}&=&0\,\,,
\end{eqnarray}
where $\mu^2\equiv -d^2_{\sigma}V_D|_{\sigma=\sigma_{\rm top}}$, which
corresponds to two different scenarios, still driven by the dilaton. For
initial fluctuation of $\sigma$ to the right of $\sigma_{\rm top}$ ({\em i.e.},
$\sigma_{in}>\sigma_{\rm top}$), the universe rolls down along the profile of
$U$ again, towards the configuration $\sigma=+\infty$: it undergoes an eternal
inflation with an unstable internal space. For initial fluctuation of $\sigma$
to the left of $\sigma_{\rm top}$ ({\em i.e.}, $\sigma<\sigma_{\rm top}$), the
universe starts with an exponential inflation ($a\propto
\exp(\sqrt{V_D(\sigma_{\rm top})}\,t$), followed\footnote{The approximations
introduced in the calculation of $\Psi$, as in Eq.(\ref{int4}), hold only
around $\sigma_{\rm top}$. To study the following evolution we must impose, to
the equation of motions, the initial conditions select by $\Psi$ around
$\sigma_{\rm top}$.} by a Friedmannian stage, {\em i.e.}, damped oscillations
of $\sigma$ around its ground state $\sigma=0$ ($b=b_0$): the internal space is
stable. Nevertheless, this peak does not select a good inflationary solution
since the flat
region of $V_D(\sigma)$ around $\sigma_{\rm top}$ is not sufficiently wide for
quantum fluctuations to be negligible during the evolution of the dilaton
throughout the region. In fact, (\ref{int4}) describes an exponential inflation
that lasts as long as
\begin{equation}\label{dur}
t<t_{\star}\equiv\frac{3\sqrt{V_D(\sigma_{\rm top})}}{\mu^2}
\log{\frac{4V_D(\sigma_{\rm top})}{\mu^2\delta_{in}^2}}\,\,,
\end{equation}
where $\delta_{in}^2\equiv(\sigma_{\rm top}-\sigma_{in})^2$. To solve the
horizon and flatness problems, it is commonly required that the exponential
phase lasted at least 65 e--foldings. Then
\begin{equation}\label{dur2}
t_{\star}\ge\frac{65}{\sqrt{V_D(\sigma_{\rm top})}}\,\,.
\end{equation}
{}From (\ref{dur}) and (\ref{dur2}) we can see that the classical initial
condition for the dilaton must be very close to $\sigma_{\rm top}$ to obtain
sufficient inflation:
\begin{equation}\label{sufinf}
\delta_{in}^2\le\left[\frac{\mu^2}{4V_D(\sigma_{\rm top})}\right]^{-1}
\exp\left[-\frac{65\mu^2}{3V_D(\sigma_{\rm top})}\right]\,\,,
\end{equation}
where $\mu^2/V_D(\sigma_{\rm top})\simeq 12$ for large $D$.

Around $\sigma_{\rm top}$, the square of the amplitude of the
wave function (\ref{peak1}) is locally approximated by a Gaussian
\begin{equation}\label{gaussian}
|\Psi|^2\propto\exp\left[-\frac{(\sigma-\sigma_{\rm top})^2}{2\delta\sigma^2}
\right]\,\,,
\end{equation}
where we have introduced the mean square fluctuation
\begin{equation}\label{fluct}
\delta\sigma^2\equiv\frac{3}{2\mu^2}V_D^2(\sigma_{\rm top})\,\,.
\end{equation}
{}From (\ref{sufinf}) and (\ref{fluct}) it follows that
\begin{equation}\label{b0}
\frac{\delta\sigma^2}{\delta_{in}^2}>\frac{D}{6}e^{182}\gg 1
\end{equation}
for values of the internal radius consistent with observational constraints,
$b_0<10^{17}$ Planck units. Thus quantum fluctuations of the dilaton span a
much larger
range than the one of the classical initial conditions that give a good
inflation. This is just due to the fact that $V_D(\sigma)$ is not sufficiently
flat around $\sigma_{\rm top}$, {\em i.e.}, $\mu^2/V_D(\sigma_{\rm top})\gg 1$.

Halliwell \cite{QC1} gets an analogous result for a model with $D=2$, no
inflaton and a monopole compactification scheme, using Hartle--Hawking
boundary conditions. Indeed this is why an inflaton must be introduced: the
dilaton alone does not lead to realistic inflationary scenarios.
The initial conditions of
slow--rolling of the inflaton in the $\phi$--direction, along the profile
of $U$, inside the stability channel are not selected by any of the amplitude
peaks of the oscillating wave function.

\subsection{\bf The semiclassical cutoff}
\label{sec:iiic}

According to Vilenkin \cite{QC1,Vile2}, the introduction of a cutoff is
required for the semiclassical approach (minisuperspace scheme, Einstein
gravity, etc.) to be valid: in fact the unbounded growth of the potential of
matter fields could take their energy density above the Planck scale.

As an example, for the 4--dimensional model with topology ${\cal R}\times
{\cal S}^3$ and a chaotic inflaton $\phi$, for the measure of the classical
solutions in the minisuperspace
\begin{equation}\label{dipi}
dP=e^{-2/3V(\phi)}d\phi\,\,,
\end{equation}
Vilenkin introduces a cutoff at the Planckian boundary $V(\phi)=1$. The reason
is that the measure (\ref{dipi}) grows monotonically with $|\phi|$, converging
to a finite limit. Although for $V(\phi)>1$ the semiclassical approximation
breaks down, the growing behaviour of (\ref{dipi}) for $V(\phi)<1$ suggests
that the universe nucleates most likely in the region $V(\phi)>1$. In this
region quantum corrections to the Einstein action (terms of higher order in
$R$) could become relevant and reverse the growth of the measure at high
energies \cite{Vile4}.

We apply an analogous prescription to the potential of matter  fields
$U(\sigma,\phi)$ in our model, in the region of the minisuperspace where $\Psi$
oscillates (the semiclassical cutoff $U=1$ takes away the whole region of the
$\sigma-\phi$ plane where $\sigma$ is negative, but large in modulus: however
here $\Psi$ is monotonic). For small $\sigma$ ({\em i.e.}, $|\sigma|\ll
\sigma_D/D$), the cutoff  takes away the subregion $V(\phi)>1$ of the region of
large $|\phi|$. In the region of small $|\phi|$ the peak of $\Psi$ around
$\sigma=\sigma_{\rm top}$ is below the cutoff only for $b_0\gg \sqrt{D}$. For
$\sigma$ sufficiently large ({\em i.e.}, $\sigma\gg \sigma_D/D$), the
semiclassical cutoff takes away only the section $V(\phi)>\exp(D\sigma
/\sigma_D)$ of the region of large $|\phi|$, while the region
\begin{equation}\label{strong}
\exp(D\sigma/\sigma_D)>V(\phi)\gg K\quad,\qquad  \sigma\gg \sigma_D/D
\end{equation}
is still below the cutoff, for any value of $b_0$.

As a result
the strong peak (\ref{tweenpeaks!}) of $\Psi$ still selects the large values
of $|\phi|$, for the classical universe in the region (\ref{strong}),
for any $b_0$, even after the cutoff $U=1$
has been introduced.
This happens because $|\Psi|^2$ depends locally only on $V(\phi)$, not
on $U(\sigma,\phi)$, unlike in the region of small $|\phi|$, where
$|\Psi|^2$ depends on $U(\sigma,\phi)\simeq
V_D$. It is easy to understand why, generalizing the results in
Ref.~\cite{Vile2} to an $n$--dimensional minisuperspace.
In a problem of superpotential barrier like (\ref{wdw}) and (\ref{sup0}),
the outgoing mode of the
Vilenkin wave function depends on the potential of the matter fields through
$|\Psi|^2=\exp(-2/3U)$, only where {\em all}
 the kinetic contributions of the matter
fields are negligible, {\em i.e.}, only where their potential $U$ satisfies
the conditions of sufficient flatness:
\begin{equation}\label{flat}
\left|\frac{\partial U}{\partial \phi_j}\right|\ll
{\rm max}\left\{U,1/a^2\right\}
\,\,,
\end{equation}
for all $j$'s ({\em i.e.} in each matter field direction). This condition is
satisfied in the strip $\sigma\simeq\sigma_{\rm top}$ of the region of small
$|\phi|$, but not in the region of large $|\phi|$, where, for $|\phi|\gg 1$,
the potential is sufficiently flat only in the $\phi$ direction.

Thus in the strip $\sigma\simeq\sigma_{\rm top}$ of the region of small
$|\phi|$, the semiclassical approximation which we used to find the peak of the
wave function (\ref{peak1}), does not hold anymore for $U\simeq V_D(\sigma)>1$,
{\em i.e.}, for $b_0<\sqrt{D}$. These are the values for which this peak
becomes comparable to the peak in large $|\phi|$. Since $V_D(\sigma)$ acts like
an effective cosmological constant for the classical solutions selected by Eqs.
(\ref{peak1}) and (\ref{int4}) we expect that quantum corrections to the
Einstein action decrease the probability of local nucleation at high energies
\cite{Vile4}. The peak of $\Psi$ at large $|\phi|$ could then remain dominant
also for $b_0<\sqrt{D}$. We remind that the peak at large $|\phi|$ corresponds
to inflationary solutions.

\subsection{\bf Conclusions}
\label{sec:iiid}
When Vilenkin
boundary conditions are applied to solving the Wheleer-DeWitt equation,
quantum cosmology  predicts a
nucleation from nothing of a classical universe in inflationary evolution
\cite{Vile2}. For this reason we have extended the analysis to an
$(N=4+D)$--dimensional model with topology ${\cal R}\times{\cal S}^3\times{\cal
S}^D$ in the equivalent ``Casimir $\sigma$--dilaton plus chaotic
$\phi$--inflaton'' scheme of Einstein gravity.

In this case, the minisuperspace is 3-dimensional and
the tunnelling analogy is no longer valid.
Nevertheless, Vilenkin's boundary conditions can be generalized in terms
that are independent of the tunnelling picture and in this form they can
be applied to our problem too. We have obtained a patchwork of approximate
analytic solutions of the Wheleer-DeWitt equation. The patchwork covers
all the relevant regions of the minisuperspace so that we have a complete
picture of the behaviour of the Universe in this model. Once  the solution
is known, information are extracted on the basis of the following criterion:
where the wave-function is
oscillatory rather than exponential, there the Universe starts its
classical evolution.

A first consideration is in order.
In 2-dimensional minisuperspace models, this criterion
leads to the conclusion that the classical evolution starts at the top (i.e.
the highest point allowed by the semiclassical cut-off) of
the potential of the inflaton field. Our calculations show that this is
still
true in our 3-dimensional minisuperspace. However the same criterion
cannot be extended to the potential of
the dilaton $\sigma$. In fact, for $\sigma\to -\infty$, and
$\phi\simeq const$
$U(\sigma,\phi)\to\infty$, but the wave function results to be exponential.

Our main conclusion is that Vilenkin wave function predicts inflation,
in multidimensional cosmology too. In fact, at the classical birth of the
Universe, we find that the law of growth
of the ordinary space is $a\propto t^p$, with $p=(D+2)/D>1$ (power-law
inflation). The inflationary phase is driven by the
initially   large and almost constant value of the inflaton $\phi$.
On the contrary, no preferred initial value of the dilaton $\sigma$
is selected by the wave function, and $\dot\sigma\gg\dot\phi$. This means
that only a very small subset of the initial conditions leads to an evolution
with $\sigma$ in the channel of the potential that gives stability to the
internal space. In the large  majority of the cases, during inflation
the internal space expands. However, the growth of $\sigma$ is only
logarithmic, {\em i.e.} it is slow enough to make the value of the radius
of the internal space at the end of inflation compatible with the
known physics. In fact, defining $N\equiv\ln(a_{\rm fin}/a_{\rm in})$,
the number of e-folds of inflation, one finds from (\ref{int3}) that
\begin{equation}\label{growth}
\frac{b_{\rm fin}}{b_{\rm in}}=e^{2N/pD}=e^{2N/(D+2)}\,\,,
\end{equation}
where $b=L\exp\sigma$ is the radius of the internal space ($L$ is a constant).
Present accelerators energies, of the order of 100 GeV, exclude the existence
of an internal space at the scale of $10^{17}$ Planck units and, in fact,
for $N=60$ and $D=6$, we obtain $b_{\rm fin}/b_{\rm in}\cong 10^{6.5}\ll
10^{17}$.
It is easily checked that the maximum allowed value for
$b_{\rm fin}/b_{\rm in}$
is reached after $N=157$ e-folds, {\em i.e.} no fine-tuning of $N$
is required.\footnote{
We assumed a chaotic potential (\ref{chaopot}) for the inflaton. In
Ref.~\cite{Amend}, a model  with a $W$--shaped potential,
such that assumed in the inflationary scenarios with a phase transition,
was also studied. Since for large $|\phi|$ the $W$--shaped potential reduces
to (\ref{chaopot}), we expect analogous conclusions about the quantum
instability of the internal space. We are investigating this case too.}

Of course the problem of the smooth connection of  the
inflationary phase to the
ordinary radiation dominated, Friedmann-Robertson-Walker phase ({\em
graceful-exit problem}) is not solved in this simplified model. We remark
that, in addition, the  mechanism for the graceful exit must also provide
the stop of the growth of the internal space.

\vspace{48pt}
\centerline{\bf Acknowledgements}
M. L. thanks E. W. Kolb for helpful discussions, and
acknowledges support by the DOE and by the NASA for hospitality at Fermilab.


\newpage
\centerline{\bf Figure captions}
{\bf Fig. 1} --  The total potential for the matter fields
($\phi$--inflaton and $\sigma$--dilaton).

{\bf Fig. 2} -- The minisuperspace section of the small
$|\phi|=const$, $\sigma<0$ in which the total potential is
$U\simeq V_D(\sigma)$.
The bold line is the boundary $W=0$ of the superpotential barrier.
The dotted line is the null--curve of constant superpotential
({\em i.e.}, where the supermetric is zero). From a comparison
with Table 1, one can get the general behaviour, oscillating or monotonic,
of the wave function, far away from the $W=0$ line.

{\bf Fig. 3} -- As in Fig. 2, but now $\sigma>0$.

{\bf Fig. 4} -- The minisuperspace section of ``large'' $|\phi|$,
$\sigma=const$, in which
$U(\sigma,\phi)\simeq e^{-D\sigma/\sigma_D}\,V(\phi)$, but
$|\partial_{\sigma}U|\gg|\partial_{\phi}U|$.
The bold and the dotted lines have the same meaning that in Fig. 2.

\newpage
\begin{table}
\begin{center}
\begin{tabular}{||c||c|c||} \hline\hline
$w=$ constant  & Spacelike & Timelike \\ \hline\hline
$w<0$    & {\bf oscillatory} & monotonic\\ \hline
$w>0$    & monotonic & {\bf oscillatory}\\ \hline\hline
\end{tabular}
\end{center}
\caption{Surfaces of constant superpotential and the beheaviour
of the wavefunction, far from the surfaces $W=0$}
\end{table}


\begin{references}

\bibitem{Kolb}
E. W. Kolb and M. S. Turner, {\it The Early Universe}, Addison-Wesley,
pub. Comp., New York, 1990.

\bibitem{Casimir}
P. Candelas and S. Weinberg, {\it Nuc. Phys.} {\bf B237}, 397 (1984)

\bibitem{FR}
P. G. O. Freund and M. Rubin {\it Phys. Lett.} {\bf 97B}, 233 (1980).
Yu. A. Kubyshin, V. A. Rubakov, and I. I. Tkachev, {\it Intl. J. Mod. Phys.}
{\bf A4}, 1409 (1989).

\bibitem{Yosh}
M. Yoshimura, {\it Phys. Rev. D} {\bf 30}, 344 (1984).

\bibitem{Sokol}
L. M. Soko\l owski, {\it Class. Quantum Grav.} {\bf 6}, 59 (1989).

\bibitem{Sokgol}
L. M. Soko\l owski and Z. A. Golda, {\it Phys. Lett.} {\bf B195}, 349 (1987);

\bibitem{Okayosh}
Y. Okada and M. Yoshimura, {\it Phys. Rev. D} {\bf 33}, 2164 (1986).

\bibitem{Hall}
J. J. Halliwell, {\it Nuc. Phys.} {\bf B266}, 228 (1986).

\bibitem{Amend}
L. Amendola, E. W. Kolb, M. Litterio, and F. Occhionero {\it Phys. Rev. D}
{\bf 42}, 1944 (1990).

\bibitem{dimred}
M. J. Duff, B. E. W. Nilsson and C. N. Pope, {\em Phys. Rep.} {\bf 130}, 1
(1986).


\bibitem{QC1}
For a review and an extended bibliography on Quantum Cosmology see:
J. J. Halliwell, {\it Proc. of the Jerusalem Winter School on
Quantum Cosmology and Baby Universes}, Singapore World Scientific,
T. Piran, S. Weinberg, (1990).

\bibitem{QC2}
J. B. Hartle, and S. W. Hawking, {\it Phys. Rev.} {\bf D28}, 2960 (1983).

\bibitem{Vile1}
A. Vilenkin, {\it Phys. Rev.} {\bf D33}, 3560 (1986).

\bibitem{Vile2}
A. Vilenkin, {\it Phys. Rev.} {\bf D37}, 888 (1988).

\bibitem{Carugno}
E. Carugno, S. Capozziello, and F. Occhionero, {\it Phys. Rev.} {\bf D47},
4261 (1993).

\bibitem{Kamen}
A. O. Barvinsky, and A. Yu. Kamenshchik, {\it Phys. Lett} {\bf B332},
270 (1994).

\bibitem{Jun2}
J. Yokoyama and K. Maeda, {\it Phys. Rev.} {\bf D41}, 1047 (1990).

J. Yokoyama, K. Maeda and T. Futamase, {\it University of Tokyo preprint}
{\bf UTAP-78/88}.


\bibitem{Vile3}
S. Del Campo and  A. Vilenkin, {\it Phys. Lett.} {\bf B224}, 45 (1989).

\bibitem{Luc}
F. Lucchin and S. Matarrese, {\it Phys. Rev.} {\bf D32}, 1316 (1985).

\bibitem{Vile4}
A. Vilenkin, {\it Phys. Rev.} {\bf D32}, 2511 (1985).

\end{references}
\end{document}